# Power Angular Spectrum versus Doppler Spectrum – Measurements and Analysis

Jan M. Kelner[1], Cezary Ziółkowski[1], Michał Kryk[1], Jarosław Wojtuń[1], Leszek Nowosielski[1], Rafał Przesmycki[1], Marek Bugaj[1], Aniruddha Chandra[2], Rajeev Shukla[2], Anirban Ghosh[3], Aleš Prokeš[4], Tomáš Mikulášek[4]

[1] Institute of Communications Systems, Faculty of Electronics, Military University of Technology, Warsaw, Poland, {jan.kelner, cezary.ziolkowski, michal.kryk, jaroslaw.wojtun, leszek.nowosielski, rafal.przesmycki, marek.bugaj}@wat.edu.pl

[2] ECE Department, NIT Durgapur, Durgapur, India, niruddha.chandra@ieee.org, rs.20ec1103@phd.nitdgp.ac.in

[3] Faculty of Engineering, Niigata University, Niigata-shi, Japan, aniz.ghosh@gmail.com

[4] Department of Radio Electronics, Brno University of Technology, Brno, Czech Republic, prokes@vutbr.cz, mikulasekt@vut.cz

*Abstract*—In this paper, we present an empirical verification of the method of determining the Doppler spectrum (DS) from the power angular spectrum (PAS). Measurements were made for the frequency of 3.5 GHz, under non-line-of-sight conditions in suburban areas characteristic of a university campus. In the static scenario, the measured PAS was the basis for the determination of DSs, which were compared with the DSs measured in the mobile scenario. The obtained results show that the proposed method gives some approximation to DS determined with the classic methods used so far.

## I. Introduction

The mobile network development that has been observed for 40 years has contributed to a significant increase in their coverage and user number, extending the scope and improving the quality of provided services. The development of new standards by international bodies (including 3GPP, ITU, ETSI, and IEEE) and the implementation of new radio and network technologies are crucial factors ensuring the success of the next generation of networks. In the case of the currently introduced fifth (5G) and the emerging sixth generation (6G), the use of higher frequency ranges (millimeter and terahertz waves) and angular-selectively beamforming (along with the use of massive multiple-input-multiple-output (MIMO) technology) are considered to be one of the most important innovations in the physical layer, which contribute to increasing the network performance parameters and extending the possibilities of the provided services [1].

Propagation measurements are the basis for the performance analysis of new radio technologies, designing radio links and mobile networks. On the other hand, these measurements are fundamentals for the development of new channel models necessary to conduct a broader analysis of the designed networks and assess the possibilities of new technologies using simulation studies [2].

Propagation measurements consist in determining various parameters and transmission characteristics of real radio channels. These characteristics are closely related to the nature of propagation environments, used antenna systems, frequency ranges, and bandwidths of transmitted signals. The greatest network capacity, i.e., the density of users (or user equipment (EU)) per area unit, is found in metropolitan and urban areas. Hence, network capacity is related to the number of base stations per area unit, which is the largest in urbanized terrains. The higher density of base stations also results from the need to ensure the appropriate quality of radio signals necessary to provide services with appropriate quality metrics. This is directly related to the fact that a greater number of obstacles (i.e., mainly buildings) in propagation paths from a transmitter (TX) to receiver (RX) occur in urban areas. So, in these scenarios, non-line-of-sight (NLOS) conditions are more common compared to the line-of-sight (LOS) conditions that prevail in rural environments.

NLOS conditions are the main cause of dispersion phenomena in the received signal. The introduction of broadband systems contributed to a significant increase in dispersion, especially in the time domain. To counteract this negative phenomenon, MIMO technology, beamforming or directional antenna systems, etc. are used in 5G and 6G systems. Power delay profile (PDP), power angular spectrum (PAS), and Doppler spectrum (DS) are transmission characteristics that illustrate radio channel dispersion in time, angle of arrival (AOA), and frequency domains, respectively [3]. The RMS delay, angular, and Doppler spreads parameters are usually used for the comparative evaluation of the dispersion phenomenon, which is determined based on PDP, PAS, and DS, respectively [4].

Assuming some condition stationarity in the selected measurement area (e.g., no movement of the surrounding elements), it can be said that the determined PAS and PDP, RMS angular, and delay spreads unambiguously describe the channel dispersion in the domain of the AOA and delay [5]. In the case of DS, the dispersion in the frequency domain additionally depends on the direction of object (TX/RX) movement with respect to the TX-RX direction. However, in [6], it was shown that DS is directly related to PAS. Thus, based on the measured single PAS, the DS can be determined for a specific position and any movement direction. It allows describing the dispersion in the frequency domain in the full range of changes in the object movement

This research was funded in part by the National Science Center (NCN), Poland, grant no. 2021/43/I/ST7/03294 (MubaMilWave). For this purpose of Open Access, the author has applied a CC-BY public copyright license to any Author Accepted Manuscript (AAM) version arising from this submission.

direction based on the PAS illustrating the dispersion in the AOA domain.

The analysis presented in [6] is based on simulation studies, while in this paper, we present this issue based on empirical measurements. To the best of the authors' knowledge, this is the first analysis of this problem, which proves the originality and novelty of the presented issue. For the purposes of this verification, we conducted PAS and DS measurements on the university campus at the frequency of 3.5 GHz, i.e., in a typical sub-6 GHz band dedicated to 5G systems [7].

The rest of the paper is organized as follows. Section II presents the theoretical foundations of the relationship between PAS and DS, which are contained in [6]. The measurement scenario and test-bed are presented in Sections III and IV, respectively. The measurement results are included in Section V. Finally, a conclusion is presented.

## II. Relationship between PAS and DS

In [6], theoretical foundations and the method of DS determination based on PAS are presented. Then, based on the assumption that the PAS is modeled by Laplacian with specific parameters [8], a numerical methodology of determining DSs based on a single PAS for different directions $\alpha$ of object movement (i.e., RX) and simulation results are shown.

PAS, $P(\phi)$, can be represented by the probability density function (PDF) of AOA, $f_\phi(\phi)$, and the total power, $P_0$, of the received signal [9]

$$P(\phi) = P_0 f_\phi(\phi). \qquad (1)$$

The relation fundamentals between DS and PAS is the relationship between the Doppler shift, $f_D$, and AOA, $\phi$, [6]

$$f_D(\phi) = f_{D\max} \cos(\phi + \alpha), \qquad (2)$$

hence, the inverse function is

$$\phi(f_D) = \pm \arccos(f_D / f_{D\max}) - \alpha, \qquad (3)$$

where $f_{D\max} = v/c$ is the maximum Doppler shift, $-f_{D\max} \le f_D \le f_{D\max}$, $v$ and $c$ are the velocity of the object (e.g., RX) and light speed, respectively.

Naturally, DS, $S_D(f_D)$, is a Doppler shift function. On the other hand, considering the above formulas, DS can be viewed as an AOA function. So [6]

$$S_D(f_D(\phi)) = P(\phi(f_D)) \cdot \mathrm{J}(\phi, f_D), \qquad (4)$$

where the Jacobian $\mathrm{J}(\phi, f_D)$ between the AOA and Doppler shift is defined for $-f_{D\max} < f_D < f_{D\max}$ [6]:

$$\mathrm{J}(\phi, f_D) = \left| \frac{\mathrm{d}\phi(f_D)}{\mathrm{d}f_D} \right| = \frac{1}{\sqrt{f_{D\max}^2 - f_D^2}}. \qquad (5)$$

Hence, we can present the final relationship between DS and PAS or PDF of AOA as follows [6]:

$$S_D(f_D(\phi)) = \frac{P(\phi(f_D))}{\sqrt{f_{D\max}^2 - f_D^2}} = \frac{P_0 f_\phi(\phi(f_D))}{\sqrt{f_{D\max}^2 - f_D^2}}. \qquad (6)$$

The numerical methodology of determining DS based on PAS is described in detail in [6].

## III. Measurement Scenarios

To empirically verify the concept presented in [6], we conducted measurements on the campus of the Military University of Technology in Warsaw. The propagation environment can be classified as a suburban area with NLOS conditions. The measurements were carried out for two scenarios: static (see Fig. 1) and dynamic (see Fig. 2). Situational maps were made using the *Google Earth* application.

The PAS was determined based on measurements in the static scenario, while the DSs were determined in the mobile scenario for the four directions of mobile traffic of the receiving station. The measurements were made for the harmonic signal (to easy determining DS) for the frequency of 3.5 GHz. Additionally, the PAS verification at 10 GHz has been planned.

### A. Static Scenario

Figure 1 illustrates the location of the transmitting (TX) and receiving (RX) parts of the measuring test-bed on the university terrain. Measurements were made near the buildings of the Faculty of Electronics (i.e., no. 45, 47, and 75).

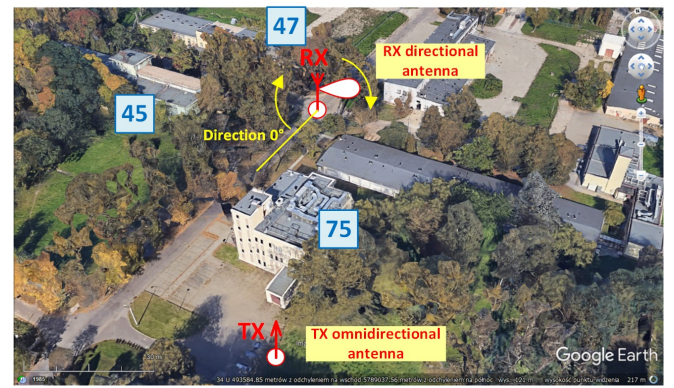

Fig. 1. Static measurement scenario.

In both scenarios, the transmitting part of the test-bed was stationary and located in the same position. The transmitting antenna was placed close to building no. 75, so as to ensure NLOS conditions at the reception site. The structure of the transmitting part is described in Section IV.*A*.

The receiving part of the measuring station was located in the middle of the road intersection, near buildings no. 45 and 47. In the static scenario, we use a directional antenna placed on a turn table. The direction of 0° for the PAS was determined along the path between buildings no. 45 and 75. The received power level was reading with a step 1°. In Section IV.*B*, the static receiving part is presented.

*B. Dynamic Scenario*

The mobile measurement scenario in the top view is illustrated in Fig. 2. In this case, the TX was located in the same place as in the static scenario. The mobile receiving part of the test-bed mounted in the vehicle moved through the intersection center near buildings no. 45 and 47.

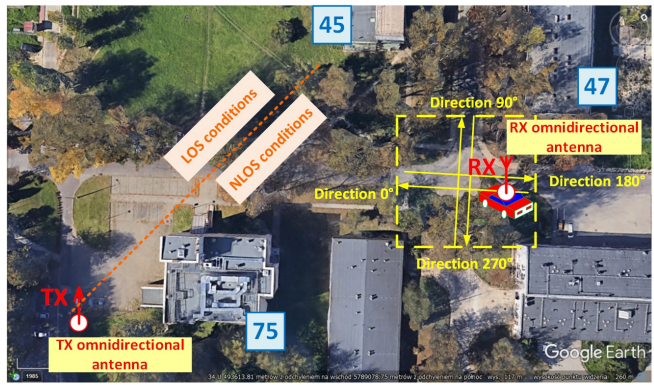

Fig. 2. Mobile measurement scenario.

Measurements involving the registration of IQ samples by the mobile RX were made for four movement directions, i.e., 0°, 90°, 180°, and 270°. The signal registration took place over a short section of the route (about 40÷50 m). The vehicle speed on the measuring section was constant and about 30 km/h. For the analyzed scenarios, the direction to the signal source (i.e., TX) relative to the direction of 0° and the intersection center (i.e., RX position for the static scenario) is about –25°. For the mobile measurement scenario, the maximum Doppler shift was approximately 97 Hz. Section IV.*C* presents the structure of the mobile receiving part.

IV. CONFIGURATION OF MEASURING TEST-BED

Section III describes two measurement scenarios that were the fundamentals for the verification of the method described in [6]. In both cases, the stationary transmitting part of test-bed was equipped with an omnidirectional antenna (see Fig. 3). In the static scenario, the receiving part was also stationary and the RX was equipped with a directional antenna mounted on the turn table (see Fig. 4). In the mobile scenario, the RX was equipped with an omnidirectional antenna that was mounted on the vehicle roof (see Fig. 5).

*A. Transmitting Part of Test-bed*

Figure 3 shows a block diagram of the transmitting test-bed part. The SMF100A signal generator by Rohde & Schwarz (R&S) generated a 3.5 GHz signal, which was additionally stabilized with an external rubidium standard. The generator was connected to the input of the transmitting antenna OMNI A0105 operating at the frequency of 3.5 GHz.

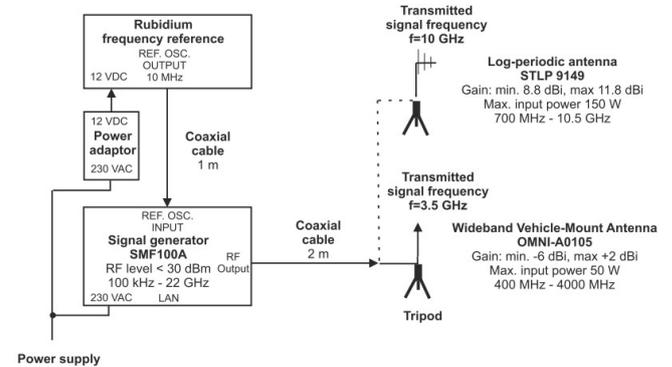

Fig. 3. Transmitting stationary part of measurement test-bed.

*B. Static Receiving Part of Test-bed*

Figure 4 depicts a block diagram of the stationary receiving part of the measuring station.

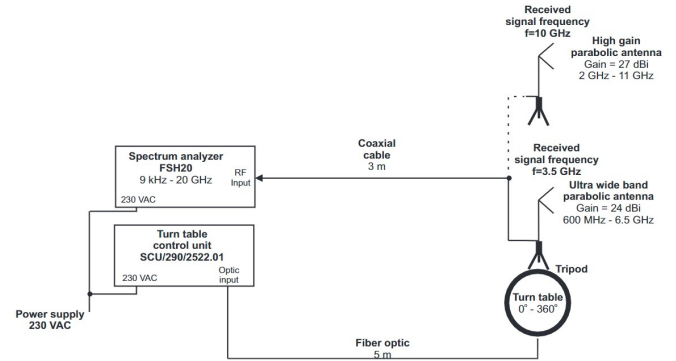

Fig. 4. Receiving stationary part of measurement test-bed.

The R&S FSH20 spectrum analyzer was connected to the inputs of two antennas:

- Ultra Wide Band Parabolic Antenna by Technical Antennas (gain 24 dBi, half-power beam width (HPBW) 8°), used to receive the signal at the frequency of 3.5 GHz,
- High Gain Parabolic Antenna by Technical Antennas (gain 27 dBi, HPBW 4°) used to receive the signal at the frequency of 10 GHz.

These antennas were mounted on the turn table to measure the received signal level as a function of the angle (azimuth) in the range from 0° to 360° with the step of 1°.

## C. Mobile Receiving Part of Test-bed

A block diagram of the mobile receiving part of the test-bed is shown in Fig. 5.

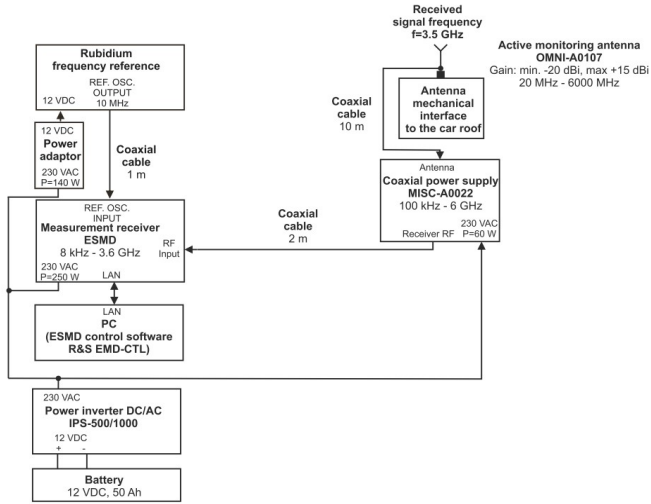

Fig. 5. Receiving mobile part of measurement test-bed.

The R&S ESMD measurement RX with an active receiving antenna, OMNI A0107, connected to its input, was used to build the mobile receiving part of the test-bed. In this case, a rubidium standard was also used as an external reference signal source to stabilize the measurement RX.

## V. EXPERIMENT RESULTS

### A. Power Angular Spectra

Based on the performed measurements of the received signal level using the test-bed for the static scenario, we determine the PAS. The measurements were carried out for the frequencies of 3.5 and 10 GHz, but for the higher frequency, the antenna HPBW is about two times lower than for 3.5 GHz. The normalized PASs obtained in the static scenario for the two analyzed frequencies are shown in Fig. 6. These characteristics are the basis for determining the DSs for the analyzed movement directions in accordance with the methodology presented in [6].

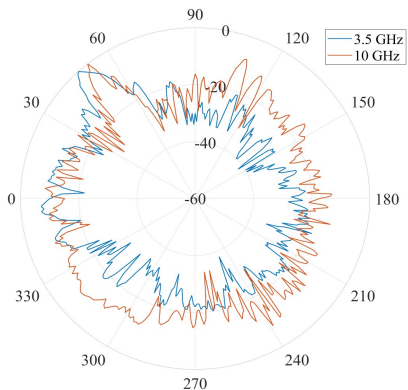

Fig. 6. Normalized PASs for 3.5 and 10 GHz.

Due to the NLOS conditions, the obtained PASs do not show a maximum in the TX-RX direction (i.e., for $\phi = 335° = –25°$). The highest signal level occurs for the direction of about 48° and 52° for 3.5 and 10 GHz, respectively. This direction is related to building no. 45, from where the main scattering of the transmitted signal comes to the RX. For 10 GHz, we used an antenna with a lower HPBW, so the obtained PAS extremum is more selective angularly. On the other hand, the obtained graphs show PAS differentiation versus frequency.

### B. Doppler Spectra

In the experiment, DSs were determined based on measurements performed for the mobile scenario. On the other hand, according to the method presented in [6], we also determined DSs based on the PAS for 3.5 GHz. The DSs obtained by the two considered methods and for four analyzed movement directions are shown in Fig. 7 (a)-(d).

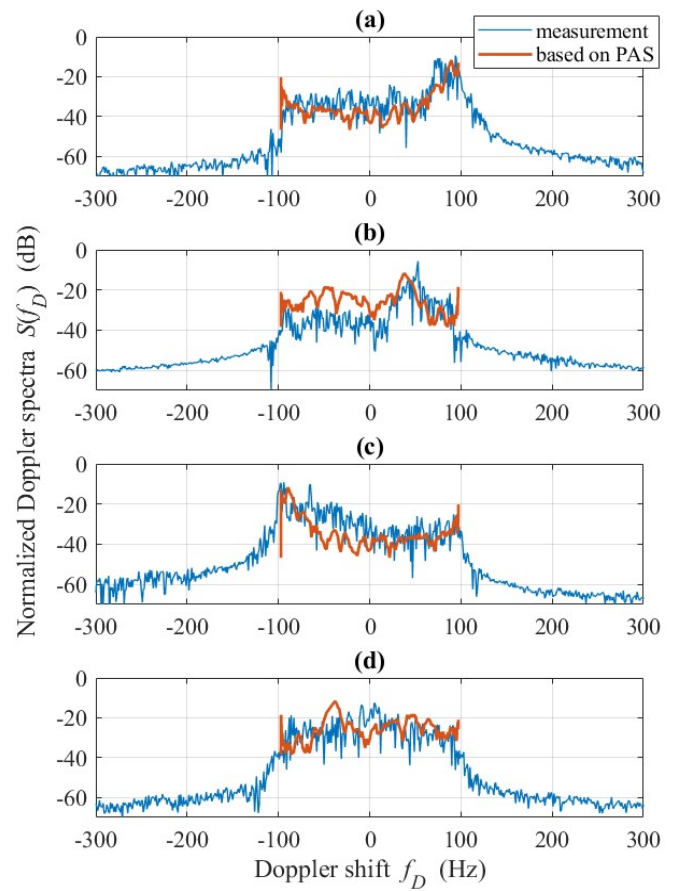

Fig. 7. Normalized DSs for selected RX movement direction: (a) 0°, (b) 90°, (c) 180°, and (d) 270°.

### C. Results Analysis

Graphs in Fig. 7 can be used for subjective comparative evaluation of DSs determined based on measurements and the method described in [6]. In this case, we see some discrepancies in the obtained channel characteristics. In the authors' opinion, the main reason for the differences is the fact that the PAS was determined pointwise, while the signal registration for DSs determination was performed spatially on the certain RX motion trajectory. Thus, the DSs obtained from the measurements are spatially averaged, while the PAS measurement was made for a specific location.

To compare the DSs obtained with the two analyzed methods, we also used two scalar measures determined for angular dispersion, i.e., average Doppler shift and RMS Doppler spread defined as follows [10][11]

$$f_{D\text{avg}} = \frac{1}{P_0} \int_{-f_{D\max}}^{f_{D\max}} f_D S_D(f_D) \mathrm{d}f_D ,\qquad(7)$$

$$\sigma_D = \sqrt{\frac{1}{P_0} \int_{-f_{D\max}}^{f_{D\max}} (f_D - f_{D\text{avg}})^2 S_D(f_D) \mathrm{d}f_D} .\qquad(8)$$

Table 1 contains these parameters calculated for all analyzed DSs.

TABLE I. TABLE STYLES

| RX movement direction | TX direction relative to RX movement direction | Average Doppler shift | | Doppler spread | |
|---|---|---|---|---|---|
| | | Based on measurements | Based on PAS | Based on measurements | Based on PAS |
| (deg) | α (deg) | $f_{D\text{avg}}$ (Hz) | $f_{D\text{avg}}$ (Hz) | $\sigma_D$ (Hz) | $\sigma_D$ (Hz) |
| 0 | –25 | 72 | 80 | 32 | 42 |
| 90 | –115 | 47 | 7 | 22 | 55 |
| 180 | 155 | –66 | –80 | 37 | 42 |
| 270 | 65 | –2 | –7 | 37 | 55 |

The largest discrepancy in results occurs for the movement direction of 90°. The best fit was obtained for the directions of 180° and 0°, and slightly worse for 270°. The results show that the method proposed in [6] can be used, e.g., in simulation studies, as a certain approximation of classical methods (i.e., those used so far).

## VI. SUMMARY

In this paper, we have made the empirical verification of the DS determination method based on PAS, which is presented in [6]. Measurements were made at 3.5 GHz in a suburban environment on the university campus. The obtained results show that the applied method can be used in simulation studies as a certain approximation of the methods of DS determination used so far. The fundamental discrepancies result from the fact that the PAS was determined pointwise, while the signal registration for DSs determination was performed spatially on a certain RX motion trajectory. Therefore, DSs based on measurements are spatially averaged.

The authors plan an additional campaign in the 10 GHz mobile scenario. Measurements with a lower-HPBW antenna (i.e., for 10 GHz), show that the obtained PAS is more accurate angularly. Moreover, to better reflect mobile measurements, the PAS should be averaged over several measurement points along the RX route. We assume that these more accurate measurements will allow for a better fit of the DSs obtained based on the two analyzed methods. If it is confirmed, we would like to highlight that method [6] allows determining the DS at a point, which is not possible in the case of mobile measurements.

ACKNOWLEDGMENT

This work was developed within a framework of the research grants: project no. 17-27068S sponsored by the Czech Science Foundation, grant no. LO1401 sponsored by the National Sustainability Program, grant MubaMilWave no. 2021/43/I/ST7/03294 sponsored by the National Science Center (NCN), grant no. UGB/22-740/2022/WAT sponsored by the Military University of Technology, and grant no. CRG/2018/000175 sponsored by SERB, DST, Government of India.